\documentclass[11pt,twoside]{article}

\usepackage{asp2006}
\usepackage{epsf}
\usepackage{lscape}
\usepackage{graphicx}
\usepackage{amssymb}

\begin{document}
\title{Implicit hydrodynamic simulations of stellar interiors}
\author{M. Viallet$^1$, I. Baraffe$^1$, C. Mulet-Marquis$^1$, E. L\'ev\^eque$^2$, R. Walder$^1$, B. Freytag$^1$, C. Winisdoerffer$^1$}
\affil{Centre de Recherche d'Astrophysique$^1$/Laboratoire de physique$^2$\\
   		Ecole Normale Sup\'erieure de Lyon\\
              46, all\'ee d'Italie, 69384 Lyon cedex 07\\
               }
              
\begin{abstract} 
We report on the development of an implicit multi-D hydrodynamic code for stellar evolution. We present two test-cases relevant for the first scientific goal of the code: the simulation of convection in pulsating stars. First results on a realistic stellar model are also presented.
\end{abstract}

\section{Introduction}

Time explicit schemes are prone to the Courant-Friedriech-Lewy (CFL) stability condition which constrains the maximum time step one can use in numerical calculations. However, in some stellar physics problems, the physical solution evolves on time-scales which are much longer than this "numerical" time-scale. This often makes calculations too cumbersome to be tractable. It is then necessary to use an implicit scheme, which does not have such a limitation on the time step. This paper describes such an implicit hydrodynamic code. The first scientific problem to be adressed with the code is the role of convection in pulsating stars. This requires an implicit approach since the typical CFL time step (for a moderate resolution) is $\Delta t_{\rm{CFL}} \sim 10$ s, whereas the convection (eddy turnover) time-scale is $\tau_{\rm conv} \sim 10^{4-5}$ s ($v_{\rm conv} \sim 0.1-0.8 c_{\rm sound}$) and typical pulsation periods are of the order of a few days ($\sim 10^{5-6}$ s) with a growth rate of the order of $\sim 10^9$ s.

\section{Code description}
\label{code}

We solve numerically the standard hydrodynamic conservation equations with radiative transfer. We use the internal energy density equation and we treat radiation transport in the diffusion limit. This approximation is suitable for optically thick region but becomes inaccurate in the photosphere and optically thin regions. Molecular viscosity is negligible in astrophysical flow, but we use artificial viscosity when robustness might be an issue, e.g. for the initial relaxation of stellar models or for shocks. The hydrodynamic equations are closed with an equation of state and opacities appropriate for the description of stellar interior structure.

We use spherical coordinates and assume azimuthal symmetry, leaving $(r,\theta)$ as independent coordinates. The spatial discretization is done on a staggered grid, using a finite volume approach. Fluxes are computed at cell interfaces with an upwind reconstruction scheme: we use either the donor cell method (first order in space) or the Van Leer method (second order in space). For the implicit temporal discretization, we use either the backward Euler scheme (first order in time) or the Crank-Nicholson scheme (second order in time). An implicit discretization results in a non-linear system which is solved by Newton-Raphson iterations. This requires the inversion of a linear system involving the jacobian matrix of the set of equations. Here the jacobian matrix is computed by numerical differencing and we use a direct solver, MUMPS (Amestoy et al. 2001), which implements an efficient and robust LU decomposition. The variable time step is chosen on empirical ground, depending on the amount of changes of variables during a time step, as described for example in Dorfi (1997).

\section{Tests cases}

For test purpose, we have also implemented in the code cartesian geometry (2D) and the equation of state of a perfect gas with a constant adiabatic index $\gamma$. Basic tests (e.g. Sedov, Sod, Barenblatt) have been successfully done for code validation during its first stages of development, so we focus here on two tests more relevant to stellar physics problems such as convection or oscillations.

\subsection{Test 1: oscillating entropy bubble}
\label{bubble}

\begin{table}[t]
\caption{Run parameters for the entropy bubble test}             
\label{table:bubble}      
\centering                          
\begin{tabular}{c c c c c}        
\hline\hline                 
Run & Time & Resolution & time step & CFL \\    
\hline                        
   1 & 50 & $50^2$ & 0.04 & 1 \\      
   2 & 50 & $50^2$ & 0.5 & 12.5 \\
   3 & 200 & $150^2$ & 0.5 & 37.5 \\
\hline                                   
\end{tabular}
\end{table}

This test is taken from Dintrans \& Brandenburg 2004 (DB in the following). We consider an isothermal stratified atmosphere with constant gravity perturbed by an entropy bubble. We compute the oscillations of this bubble around its equilibrium position and we analyze the spectrum of internal gravity and sound waves excited by the bubble. The initial setup, parameters and units normalization are similar to DB and are therefore not reproduced here. The Brunt-V\"ais\"al\"a frequency of the atmosphere is $N\sim 0.82$. The kinematic viscosity $\nu$ is set to $5\times10^{-4}$ in all runs. The thermal conductivity $k_r$ is set to zero, as we are interested in the adiabatic evolution of the system. We use a cartesian domain $(x,z) \in [-0.5,0.5]\times[0,1]$. Periodic boundary conditions are used in the horizontal directions and non penetrative conditions (zero vertical velocity) are used at the bottom and top boundaries. All the run parameters are summarized in Table \ref{table:bubble}.

As in DB, we describe each wave mode by two integers $l,n \ge 0$ which are the number of nodes in the horizontal and vertical direction. In this test we look at two types of waves (for the derivation of these properties we refer the reader to DB): 1) Vertical sound waves which have frequencies $\omega = (n+1)\pi$. These waves have $l=0$ since they have no horizontal structure. 2) Internal gravity waves which have frequencies $\omega < N$ and are characterized by $l\ne0$ (internal gravity modes cannot propagate in the vertical direction).

\begin{figure*}[t] 
   \centering
   \parbox{0.4\linewidth}{\includegraphics[width=0.7\linewidth,trim = 40 20 80 100,angle=90]{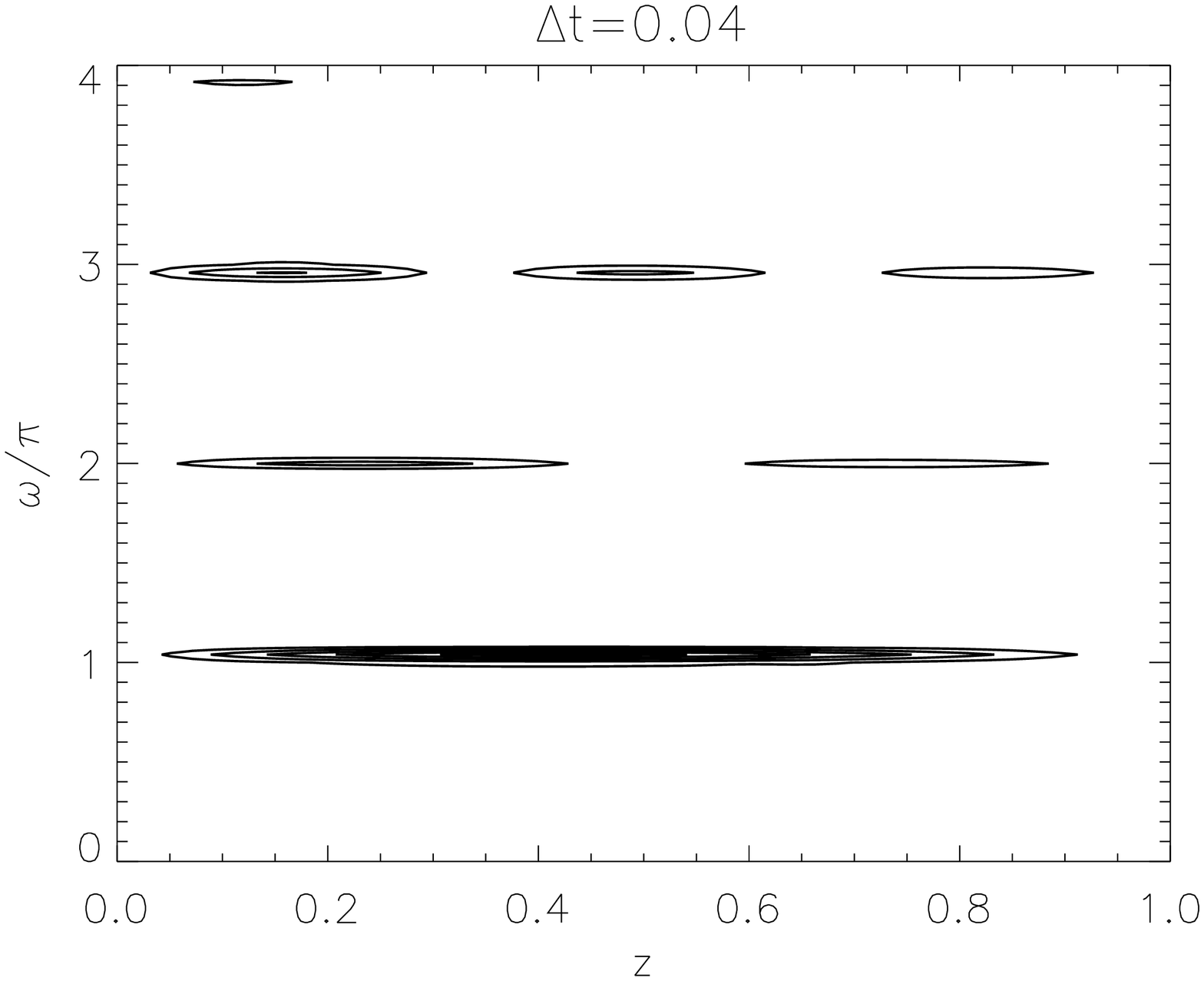}}
   \parbox{0.4\linewidth}{\includegraphics[width=0.7\linewidth,trim = 40 20 80 100,angle=90]{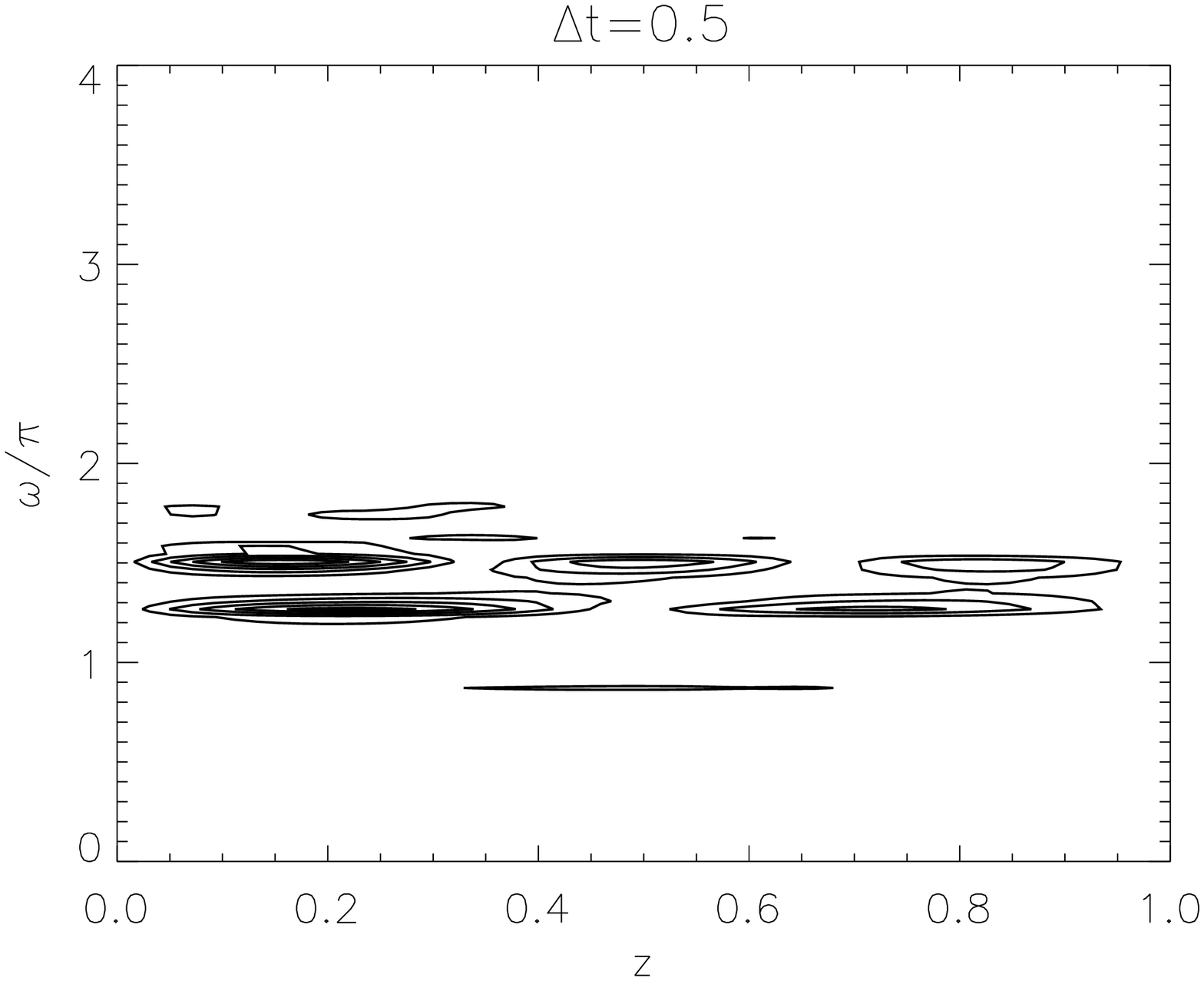}}
   \caption{Accoustic modes in the $(z,\omega)$ plane for $\Delta t = 0.04$ (left) and $\Delta t = 0.5$ (right).}
   \label{fig:bubble_acc_cfl}
\end{figure*}

These two kinds of waves occupy well separated regions of the frequency spectrum. Vertical sound waves have vertical periods $P \le 2$ and will therefore be more sensitive to the time step. Internal gravity modes have longer periods $P \gtrsim 7.3$ and will therefore be less sensitive. This test thus provides a good analysis of the influence of the time step on the accuracy of the numerical results.

We first analyze the vertical sound wave spectrum by using method 2 described in DB, with run 1 \& 2 (c.f. Table \ref{table:bubble}). In this method, the wave spectrum is represented in the $(z,\omega)$ plane. Note that for a given time step $\Delta t$, the Nyquist theorem states that no periodic signal with frequency greater than $\omega_N = \pi / \Delta t$ can be resolved. Also the temporal Fourier transform has a frequency resolution which is $\Delta \omega = 2\pi/T$, where $T$ is the duration of the simulation. The results are shown in fig. \ref{fig:bubble_acc_cfl} for two different time steps. For $\Delta t = 0.04$ we recognize the signature of the $n=0,1,2$ modes (the $n=3$ mode is below the range of the isocontour levels). Furthermore, all the modes are located at the correct frequencies (within $\Delta \omega \simeq 0.125$). For $\Delta t = 0.5$, one has $\omega_N = 2\pi$ thus normally allowing for modes $n=0$ and $n=1$ only. In the map one can recognize the $n=0,1,2$ modes. None of them is located at the correct frequency, it is also surprising to find the $n=2$ mode whose eigenfrequency is above the Nyquist frequency. In this case it is clear that the time step is too large to accurately resolve the sound waves. To analyze the gravity modes we use the method developed in DB: we project the velocity field on the anelastic eigenvectors. With this method we obtain the individual mode amplitudes $c_{ln}$ which are the coefficients in the eigenvector expansion. With such a decomposition it is straightforward to obtain the mode frequency. We have checked (not shown here) that for $\Delta t=0.5$ (run 3), the gravity mode frequencies perfectly match the analytical predictions. 

This test highlights the importance of the choice of the time step to correctly describe the physical process of interest. In the present case, sound wave amplitudes are much lower than the amplitudes of the gravity modes (the initial setup is almost in hydrostatic equilibrium) and therefore they do not play an important role in this problem, one can then safely use a larger time step (e.g. $\Delta t=0.5$) to study the internal gravity modes, as shown by the excellent agreement with predicted frequencies. Obviously one could not do that if the initial setup was intended to be far from hydrostatic equilibrium as sound waves would then be at least as important as internal gravity modes.

\subsection{Test 2: Rayleigh-B\'enard convection}
\label{convection}

\begin{table}[b]
\caption{Run parameters for the Rayleigh-B\'enard test}             
\label{table:rb}      
\centering                          
\begin{tabular}{c c c c c c}        
\hline\hline                 
Run  & $m$ & Ra & Resolution & max. time step & max. CFL \\    
\hline                        
   1 & 1 & $5\times 10^3$ & $50^2$ & 0.04 & 1 \\      
   2 & 0 & $10^4$ &$50^2$ & 0.5 & 12.5 \\
   3 & -0.9 & $10^5$ & $50^2$ & 0.5 & 37.5 \\
\hline                                   
\end{tabular}
\end{table}
Since one of the first planed astrophysical applications of our implicit code is stellar convection, the simple problem of Rayleigh-B\'enard convection is a good test. The initial setup and parameters are taken from Brandenburg et al. (2005). We consider a stratified atmosphere in a cartesian domain $[-0.5,0.5]\times[0,0.5]$. The initial atmosphere is a polytropic model characterized by a polytropic index $m$ so that initially $\rho \propto T^m$. We consider a perfect gas with $\gamma=5/3$. For $m<1/(\gamma-1)=3/2$ the temperature stratification is super-adiabatic and the atmosphere is dynamically unstable. If in addition the Rayleigh number is above its critical value, the atmosphere will be also convectively unstable. In Brandenburg et al. (2005) it is shown that for a given value of $m$, one can deduce the expected ratio of the convective flux $F_{\rm conv}$ to the total flux $F_{\rm tot}$. For example, for $m=$1, 0 and $-0.9$, this ratio is respectively $20\%$, $60\%$ and $96\%$. The purpose of this test is to check if we can reproduce these values.

\begin{figure*}[t] 
   \centering
   \parbox{0.32\linewidth}{\includegraphics[angle=90,width=1.1\linewidth]{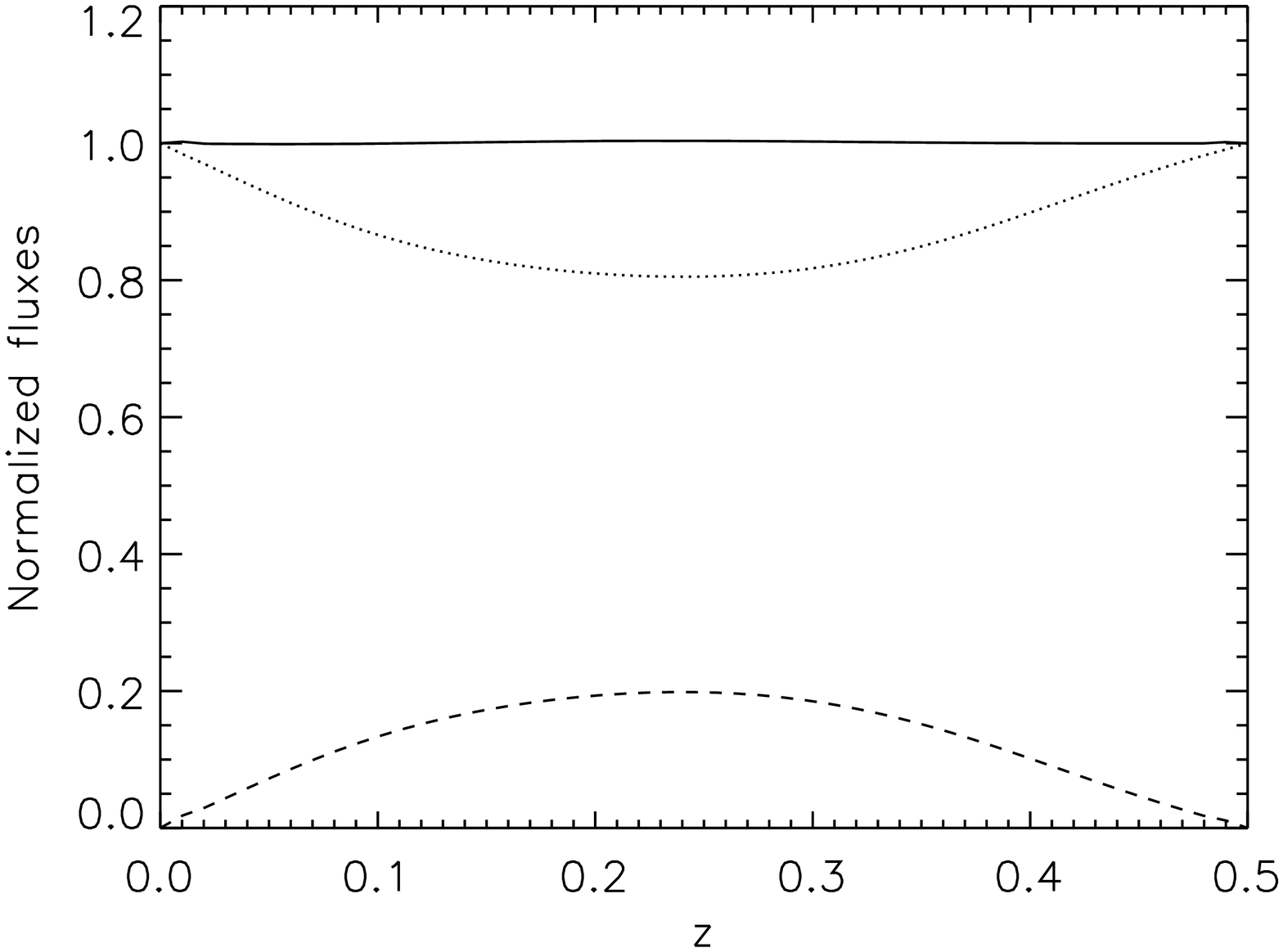}}
   \parbox{0.32\linewidth}{\includegraphics[angle=90,width=1.1\linewidth]{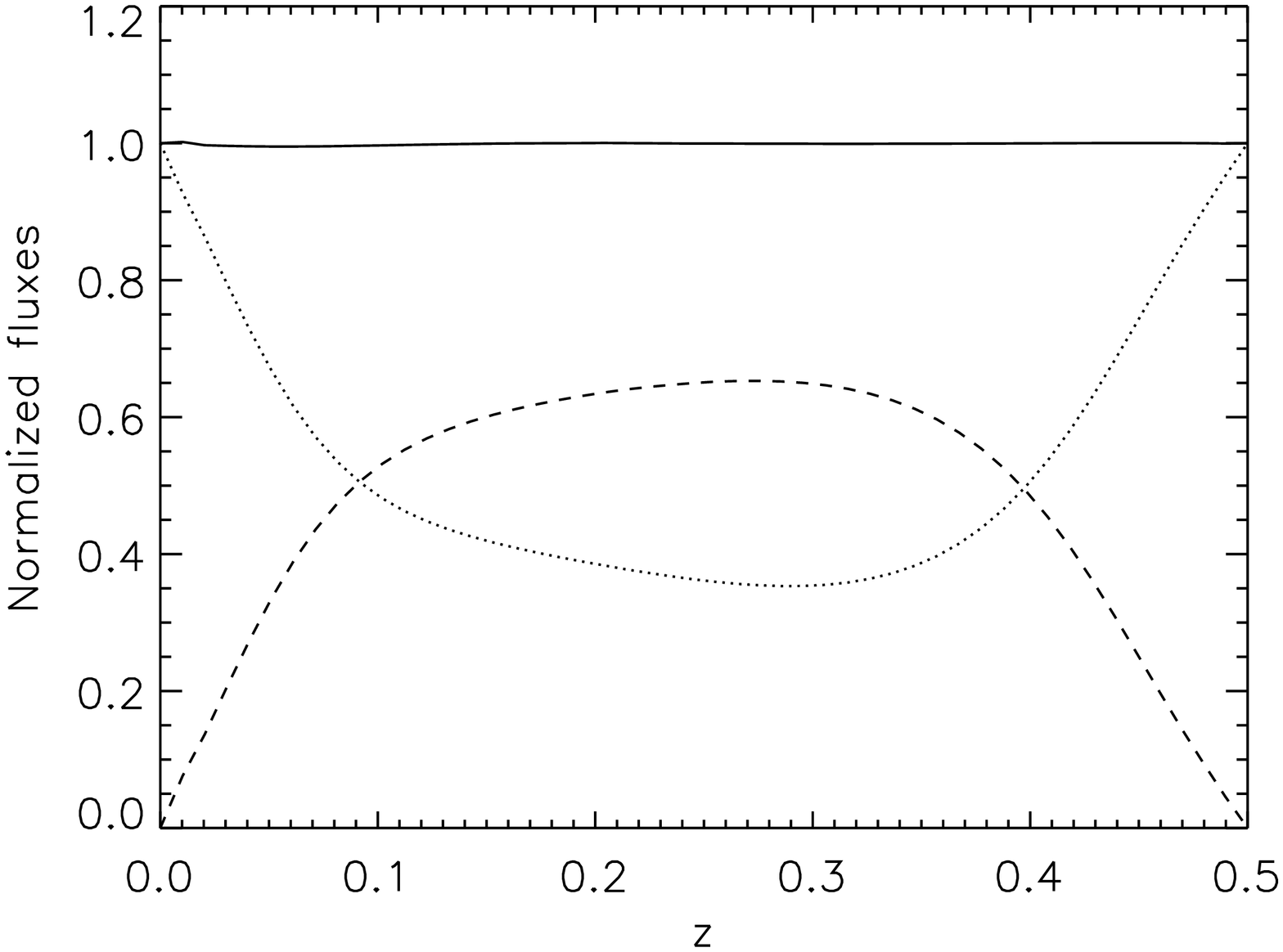}}
   \parbox{0.32\linewidth}{\includegraphics[angle=90,width=1.1\linewidth]{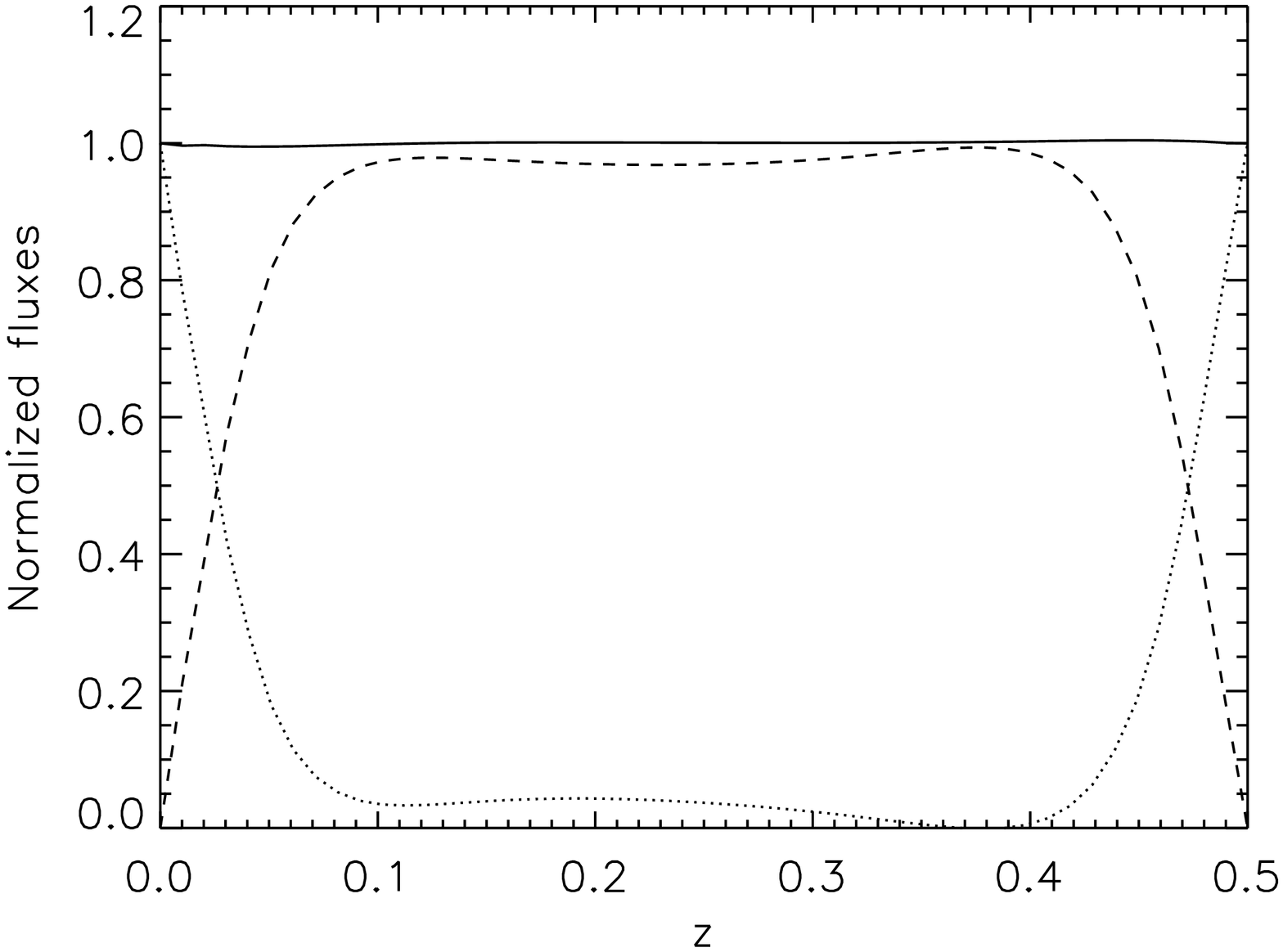}}
   \caption{Vertical profiles of the horizontally averaged fluxes (full: total flux, dashed: convective flux, dotted: radiative flux) for $m=1,0,-0.9$ (from the left to the right). The expected ratio of $F_\mathrm{conv}/F_\mathrm{tot}$ are 20\%, 60\% and 96\%.}
   \label{fig:rb}
\end{figure*}

We use a constant viscosity $\nu = 5\times 10^{-4}$ and a constant thermal conductivity $k = 10^{-3}$. Periodic boundary conditions are used in the horizontal direction. Non penetrative conditions are used at the top and at the bottom where in addition the total flux $F_\mathrm{tot}$ is imposed. The Rayleigh number is computed as $\mathrm{Ra} = \frac{1}{T_0} \frac{dT}{dz} \frac{g d^4}{\bar \chi \nu}$ where  $\bar \chi = k/(\rho_0 c_v)$ is a mean value of the thermal diffusivity ($T_0$ and $\rho_0$ are mean values of the temperature and density). All simulations are done on a $50^2$ grid. All run parameters are summarized in Table \ref{fig:rb}. In each run the system converges to a steady state showing two convective cells. This is not surprising since the aspect ratio of our domain is chosen so that the most unstable mode almost exactly fits in the box. This steady state is reached after about 100 units of time. In practice the time step converges at some value shown in Table \ref{table:rb}. The $m=-0.9$ case has a smaller time step, which is not surprising since it is the most unstable case investigated here.

We define the following horizontally averaged fluxes $F_\mathrm{rad}=< -k \frac{d T}{dz} >$, $F_\mathrm{enth} = < \rho u_z c_p (T - <T> ) >$, $F_\mathrm{kin} = < \rho u_z \frac{u^2}{2} >$ and $F_\mathrm{visc} = < - \vec u . \bar{\bar{\tau}}>$, where $< . >$ stands for horizontal averaging. In the steady state, their sum should be equal to $F_\mathrm{tot}$. In practice we find that this is the case within $0.5\%$. The kinetic energy flux $F_\mathrm{kin}$ and the viscous flux $F_\mathrm{visc}$ are found to transport less than $\sim 1\%$ of the total flux so that $F_\mathrm{conv}\sim F_\mathrm{enth}$. Figure \ref{fig:rb} displays the vertical profiles of these fluxes and shows that the expected ratios are recovered. Note that in the last case, almost all the flux is transported by convection.

\begin{figure*}[t] 
   \centering
   \includegraphics[width=3.3cm,angle=90,bb=50 20 228 678]{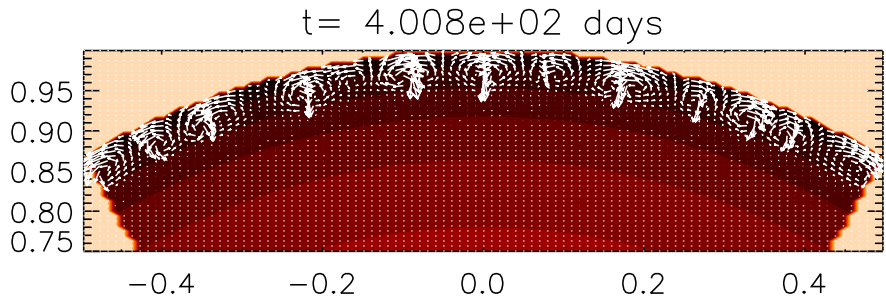} 
   \caption{Snapshot showing the convective eddies. The axis are normalized by the stellar radius $R_\star$. The color codes the logarithm of the density (light to dark color means high to low density).}
   \label{fig:conv_sn}
\end{figure*}

\section{Preliminary results under realistic stellar interior conditions}

We consider a hot, almost fully radiative star model to test the ability of the code to describe convection in stellar conditions. The initial stellar model used for the hydrodynamic calculations is produced by a 1D stellar structure code and has $T_\mathrm{eff}=7500$ K, $M=5M_\odot$, $\log(L/L_\odot)=3$ and $\log(g)=2.6$. The mixing length parameter (hereafter MLT) is taken as 1.7. This model lies outside the instability strip of Cepheids so it is stable against pulsations. This model has a small convective region near the surface, in the H/He ionization region around $T\sim10^4$~K. Our initial model covers $60\%$ of the stellar radius $R_\star \sim 19R_\odot$.

We consider a numerical domain $(r,\theta) \in [0.4R_\star,R_\star]\times[\pi/3,2\pi/3]$. We use a restricted angular domain to increase the angular resolution without sacrifying too much CPU time. The grid size is $150^2$. We use reflective boundary conditions in the $\theta$ direction and non-penetrative conditions at the top and at the bottom of the envelope. An energy flux consistent with the stellar luminosity is imposed at the bottom. The boundary conditions for the energy flux at the outer boundary are not straightforward and different solutions are possible. As a preliminary solution, we impose that the last layer of cells radiates toward the exterior a flux equal to $\sigma T^4$, where $T$ is the temperature of the cells. This is justified by the fact that the outer boundary conditions of the initial 1D model relies on the Eddington approximation and is defined at the photosphere, at an optical depth $\tau_{\rm Ross}=2/3$ where $T=T_{\rm eff}$.

The initial 1D model does not exactly correspond to an equilibrium state of the hydrodynamic code. Consequently, the model undergoes an initial phase of relaxation, for which we find more convenient to use a dissipative, more robust, 1st order scheme (Euler backward + donor cell). This initial relaxation phase lasts for about 100 days during which the model reaches hydrostatic and thermal equilibrium, which corresponds to the Kelvin-Helmholtz time-scale of the envelope. Convective eddies then develop near the surface (see Fig. \ref{fig:conv_sn}). At this stage, we switch to a second order scheme. The eddies remain throughout the whole simulation, which lasts over 700 days. The mean time step is $\sim 10^4$ s which corresponds to a CFL number of $400$. 

The eddies pattern show a non-steady behavior, but always remain confined near the top boundary. In table \ref{table:comp}, it is shown that our results are consistent with MLT. It is important to note that the mean time step of the simulation is lower by a factor of ten than the maximum eddies turnover time-scale. This tells us that the latter quantity is certainly relevant to determine the time step in order to accurately describe convection.

\begin{table}[t]
\caption{Comparison with MLT.}          
\label{table:comp}      
\centering                          
\begin{tabular}{c c c}        
\hline\hline                 
   & Simulation & MLT\\
\hline                        
 Convection Mach number $v_\mathrm{conv}/c_s$ & 0.02 & 0.05 \\ 
 Eddy turnover time-scale  & $10^5$s - $10^7$ s & $10^5$ s\\ 
 Eddy size & $\sim 5 H_p$ & 1.7 $H_p$\\ 
 $F_\mathrm{conv}/F_\mathrm{tot}$  & $3\times 10^{-4}$ & $2\times 10^{-3}$\\
\hline                                   
\end{tabular}
\end{table}

\section{Conclusion}

We have presented a new multi-D hydrodynamic implicit code. As shown in section \ref{bubble}, high CFL simulations are efficient to study physical processes proceeding on time-scale larger than the CFL limit. One should always consider numerical results with care, as an inappropriate time step can give significantly wrong results. We have shown that the code is able to handle strong convection, at least in a simple setup. Our first results on a realistic (weakly convective) stellar model are encouraging and provide very promising perspectives for the first multi-D  simulations of convection in pulsating star, a challenge in the field of asteroseismology.

\acknowledgements This work was funded by the Agence Nationale de la Recherche under the ANR project number NT05-3 42319 (Star multi-D) and by the Programme National de Physique Stellaire (PNPS).

\end{document}